# OBSERVATION OF THE MARCH MAXIMUM IN THE DAEMON FLUX FROM NEOs IN THE YEAR 2005: NEW EFFORTS AND NEW EFFECTS

E. M. DROBYSHEVSKI

*Ioffe Phisico-Technical Institute, Russian Academy of Sciences,194021 St-Petersburg, Russia*
*E-mail:* emdrob@mail.ioffe.ru

The experiments of 2005 aimed at detection of low-velocity (~10-15 km s$^{-1}$) daemons falling on to the Earth's surface from Near-Earth, Almost Circular Heliocentric Orbits (NEACHOs) have corroborated once more the existence of the March maximum in their flux by raising its confidence level to 99.99%. In addition, these experiments permitted us to identify several FEU-167-1-type PM tubes, with a few times thicker inner Al coating, which appear to be capable to detect, without any scintillator, the crossing of negatively charged daemons. As a result, detection efficiency increases tens of times, thus raising the measured level of the March daemon flux to $f_\oplus \geq 0.5 \times 10^{-7}$ cm$^{-2}$s$^{-1}$.

## 1  DAEMONS AND DETECTION OF THEIR NEAR-EARTH POPULATION. ON UNCONTROLLABLE PARAMETERS IN THER DETECTION

Dark Electric Matter Objects, daemons, presumed to be relic Planckian elementary black holes with $M \approx 3 \times 10^{-5}$ g and $r_g \approx 2 \times 10^{-33}$ cm, carry an electric charge of up to $Ze \sim G^{1/2}M \sim 10e$ whose repulsion is compensated by self-gravitation (e.g., Markov, 1966; Barrow, Copeland and Liddle, 1992; Ivanov, Naselsky and Novikov, 1994; Drobyshevski, 1996). When moving with velocities on the astronomic scale (~10-100 km s$^{-1}$), they do not trigger a scintillator (Markov, 1966).

Nonetheless, in our *scintillation-based* (sic!) experiments we drew upon (Drobyshevski, 2002):
(1) The capability of *negative* daemons to capture atomic nuclei in matter with ensuing release of binding energy $W \approx 1.8ZZ_nA^{-1/3}$ MeV (~100 MeV) and the resultant emission of atomic electrons and ejection of nucleons from the nuclei;
(2) Successive daemon-stimulated *proton decays*, with ~1 μs mean interval, inside the remainder of the captured nucleus (Drobyshevski, 2000). The latter process lowers the charge of the daemon/nuclear remainder complex down to $Z_{eff} = |Z_n| - |Z| < 0$, thus making possible capture by the daemon of a new nucleus. If the daemon propagates in matter, the process repeats. Thus, in contrast to conventional nuclear physics particles, daemons do not produce a continuous track in a scintillator. Their path is recorded by events which are spaced fairly far both in time and space.
(3) The flux of daemons as a part of the Galactic disc or halo population is fairly weak, about ~$3 \times 10^{-12}$ cm$^{-2}$s$^{-1}$. This is why we focused our attention in the experiments on detection of slow ($V \sim$ 10-30 km s$^{-1}$) objects captured from the Galactic disc low-velocity population (Bahcall, Flynn and Gould, 1992) by the combined action of the Sun and the Earth and building up in heliocentric orbits crossing the Earth's orbit. In their motion with such a low velocity through condensed matter, daemons capture at $Z_{eff} = -1$ a nucleus in a distance of ~1-10 µm. Rough estimates suggest that their flux intercepted by the Earth may reach as high as $f_\oplus \sim 3\times10^{-7}$ cm$^{-2}$ s$^{-1}$ (Drobyshevski, 1997).

We started experiments in autumn of 1996 and, after exploring several avenues, our choice fell on the simplest possible detection system (Drobyshevski, 2002).

The detector consisted of two transparent polystyrene plates, 4 mm thick and 0.5×0.5 m$^2$ in size, which were coated on the underside with a 3.5-mg cm$^{-2}$-thick (~8 µm) layer of ZnS(Ag) powder. Heavy nonrelativistic particles, like α-particles or protons, release in crossing such a layer ~1 MeV of energy, whereas β and γ rays and cosmic ray muons, <10 keV. The plates separated by a 7 mg cm$^{-2}$ sheet of black paper were arranged horizontally at a distance of 7 cm from one another at the center of a tin-iron (0.3 mm Fe + 2 µm Sn on both sides) cubic box 51 cm on a side. They were viewed each by its FEU-167 PM tube with photo-cathodes mounted flush with the horizontal sides of the box. The top side of the box was made of two sheets of black paper. The PM tubes were powered, according to the specifications, by a voltage corresponding to their sensitivity of 10 A lm$^{-1}$ (actually, they were adjusted with a photodiode to operate at the same sensitivity close to this value).

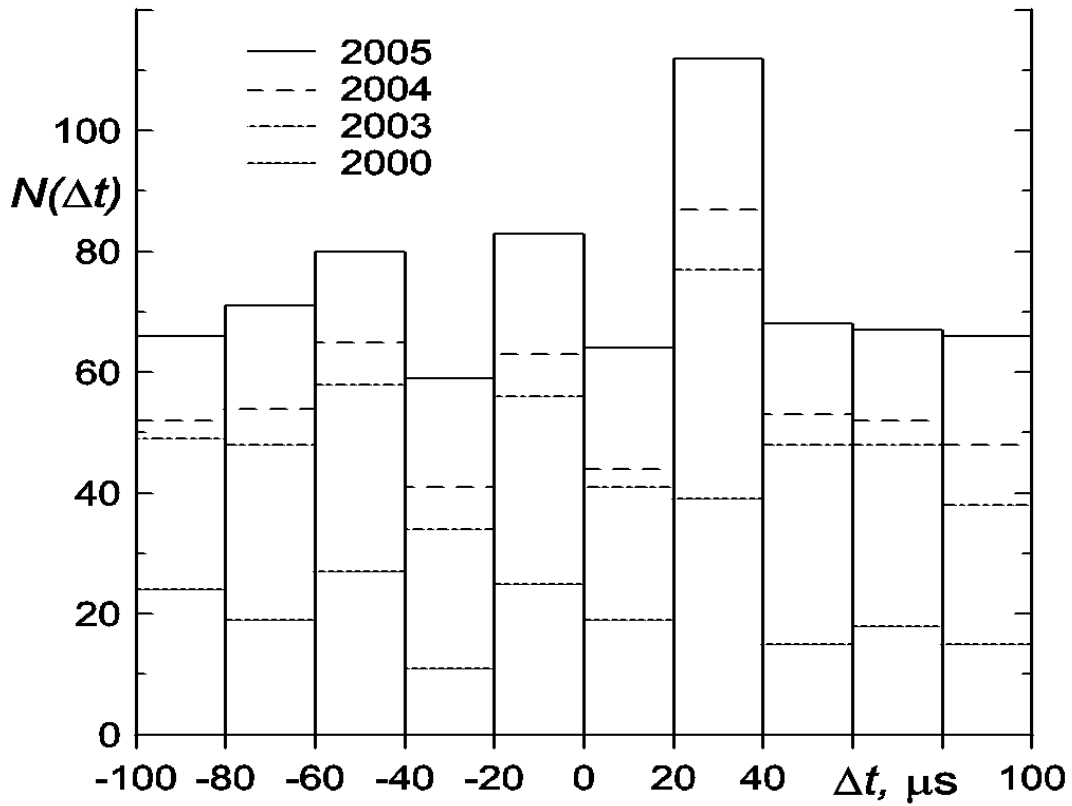

FIGURE 1. The sum of $N(\Delta t)$ distributions for the system of four modules, #1, #2, #3, and #4, collected during the months of March: 2000 (24 February-27 March; $N(\Delta t)$ for the interval $-100 < \Delta t < +100$ µs: 24, 19, 27, 11, 25, 19, 39, 15, 18, 15), 2003 (24 February – 27 March; 25, 29, 31, 23, 31, 22, 38, 33, 30, 23), 2004 (8-20 March; 3, 6, 7, 7, 7, 3, 10, 5, 4, 10), and 2005 (three modules #1, #2 and #24; 5-25 March; 14, 17, 15, 18, 20, 20, 25, 15, 15, 18). The 30-µs maximum in the total distribution exceeds the mean level by 33 events and has a significance of 3.63$\sigma$ (C.L. > 99.97%) (see also footnotes [1] and [6]).

The outputs of both PM tubes were supplied to an S9-8 double-trace digital storage oscillograph. The operation was triggered by a signal from the top PM tube (the first oscillograph trace). If a signal was detected on the second trace within ±100 µs from the beginning of the first one, the event representing two digitized oscillograms was stored in computer memory. Two types of events were observed; one of them, with a flat maximum at ~2.5 µs, is characteristic of α-particle scintillations (referred to in what follows as Heavy Particle Scintillations, HPSs), and the other, with a sharp maximum at 1-1.5 µs, is typically assigned to passage of cosmic rays or PM tube background, including signals produced in the PM tube by cosmic rays (NLSs, or Noise-Like Signals).

We started by building four detection modules of the above type.

The experiments performed on them in March 2000 revealed a clearly pronounced maximum at +20 < $\Delta t$ < 40 µs in the $N(\Delta t)$ distribution of events in the time displacement $\Delta t$ of the beginning of the signal on the second trace relative to the beginning of the HPS signal on the first trace (see Fig. 1 here, as well and Fig. 2 in Drobyshevski, 2002). We assigned this event to the crossing of NEACHO daemons with $V \sim$ 10-15 km s$^{-1}$ (Drobyshevski *et al.*, 2003a). This is argued for, if by nothing else, by the small width of this 30-µs peak. Indeed, the velocity excess of NEACHO objects compared with the Earth's orbital velocity of 29.8 km s$^{-1}$ may lie within 0 < $|\Delta V|$ < 12.3 km s$^{-1}$ (objects with $|\Delta V| \rightarrow$ 12.3 km s$^{-1}$ should move in other than NEACHOs). A NEACHO object enters the detector with a velocity $V_{fall}$ = (11.2$^2$ + $\Delta V^2$)$^{1/2}$, i.e., 11.2 ≤ $V_{fall}$ < 16.6 km s$^{-1}$. The daemon path length in the detector, depending on the actual inclination of its trajectory, varies from 29 to 42 cm (see below). Therefore, the 20-µs width of this peak should be identified with objects moving with velocities 7.5 < $V_{fall}$ < 22 km s$^{-1}$. We readily see that this interval encompasses the possible values of $V_{fall}$ from NEACHOs with a large margin.

The above emphasizes the importance of detection and study of the 30-µs peak. Observations carried out during the subsequent months, which involved also variation of the parameters of the system (for instance, introduction of tinned iron sheets between the scintillators, turning the modules upside down etc.) did not, however, yield reproducible results while did not at the same time reveal any errors in operation of the system. Starting with October 2000, one more four-module system (modules #21-#24), which could be operated in somewhat different regimes, was assembled. Its top polystyrene plates coated on the underside by ZnS(Ag) were 1 mm thick. The experiment revealed a slightly lower sensitivity of the new compared to the old system. In June 2002, both systems were transferred to an air-conditioned room.

The lack of month-by-month reproducibility of the data amassed from March 2000 to March 2001 suggested the existence of a seasonal variation in the daemon flux (Drobyshevski *et al.*, 2003a), a conjecture borne out by a continuous exposure of the second four-module system (without variation of its parameters), first from April 2001 through April 2002 (Drobyshevski *et al.*, 2003b), then to April 2003 (Drobyshevski, 2005b), and (with a break) to December 2004 (Drobyshevski, 2005a). This permitted us to detect at a confidence level of 99.9% a variation in parameters of the near-Earth daemon flux with a period $P$ = 0.5 yr and with maxima in February-March and again in August-September. We believe that about this time Earth crosses the areas with crowding NEACHOs for objects captured into these orbits by the combined action of the Earth and the Sun as a result of motion of the Solar system relative to the daemon population of the Galactic disk. This is where the probability for the daemons to transfer into the Earth-crossing orbits (and further to the NEACHOs) following their crossing and slowing down in the Sun reaches a maximum, because the projection of the Earth's orbital velocity on the Sun's apex direction achieves in these areas its largest value (Drobyshevski, 2005b).

On the other hand, we have not succeeded in observing a maximum in the March HPS $N(\Delta t)$ distributions for $\Delta t \approx$ +30 µs in subsequent years which would be as significant as the one in the year 2000. Fig. 1 presents also the March data summed over the 2000, 2003, and 2004 years. While the significance of the 30-µs peak increases from one year to another to reach 3.33$\sigma$ in 2004 (C.L. ≈ 99.9%)[1], these results cannot be accepted as satisfactory. Indeed, it is easy to verify that a simple repetition of the results of March 2000 would have yielded a significance of ≈4$\sigma$ for the 30-µs maximum.

We are going to describe in Section 3 below one more attempt to reproduce our first (and, thus far, the best) results of the year 2000. This attempt was crowned only by a partial but very essential success. We have understood that among such uncontrollable factors as, for instance, the use of PM tubes producing inevitable noise because of radioactive potassium present in their glass envelopes, may be the response of a PM tube to a daemon crossing it. This response is caused by deviation of the thin internal structural components of a PM tube from specified values, which may originate from unavoidable variations in the technology of their fabrication. Such deviations in no way affect the photometric characteristics of the device specified in its certificate. As this will be shown in Section 5, however, it is because of such deviations that some PM tubes of the FEU-167-1 series can themselves serve as sensitive detectors of daemons crossing them.

## 2 FACTORS INFLUENCING DETECTION SENSITIVITY. ADVANTAGES OF AN EVACUATED DETECTOR

The reasons responsible for the drop in detector efficiency after March 2000 have been constantly stirring our concern for the following five years. The only natural way to reveal them seemed to purposefully vary the parameters of the system that, judged from the standpoint of our hypothesis of the existence and probable properties of negative daemons, could exert any influence. Obviously enough, the situation is complicated by the fact that we are not able yet to identify unambiguously the events initiated by the passage of daemons and, thus, have to use statistical data as a basis, where detector crossing by daemons falling from NEACHOs with $V \sim 10\text{-}15$ km s$^{-1}$ produce a 30-µs maximum in the March $N(\Delta t)$ distribution. If we had not had this maximum, it would have been much more difficult to insist on the existence of daemons.

Unfortunately, the data obtained with detectors with modified parameters can be compared only once a year. Nevertheless, we have been making these attempts during the recent years. It was found, for instance, that the scintillator layer degrades with time because of the finest fraction in the ZnS(Ag) powder falling out. We checked the conjectures that daemons become poisoned by Al or Si nuclei in passing through paper containing kaolin $Al_2Si_2O_5(OH)_4$ (in addition to black paper, we used sometimes white one to wrap up the black paper in order to increase the reflection of light), that absolute air humidity in the room may also contribute, etc.

---

[1] We calculate the statistical significance of the maximum in the simplest way possible, i.e., divide its excess over the arithmetic mean of the number of events per bin in the distribution by the root square of the number of events in the bin with the maximum. This is a fairly moderate estimate. Indeed, one can find the excess of the maximum over a weighted mean for all ten bins (which is less than the arithmetic mean) and divide it by the square root of the sum of squared errors, i.e., by the square root of the sum of the number of events in the maximum and the squared error of the weighted mean (see, e.g., Taylor, 1982). One could also calculate the weighted mean for nine bins (i.e., excepting the bin with the maximum), find the excess of the maximum over this weighted mean, and again express it in terms of the squared sum of errors. One can readily verify that for the March 30-µs peak in the $N(\Delta t)$ total distribution for the years 2000, 2003, 2004, and 2005 (Fig.1), being calculated by this way, these values for statistical significance become 3.63$\sigma$ (C.L. = 99.97%), 3.72$\sigma$ (99.98%), and 3.96$\sigma$ (99.99%). One could, of course, continue refinement of the estimate further, but this would hardly add anything of value to the experiment as such.

The latter consideration drew from a basically simple idea that detector efficiency is proportional to the time during which a daemon resides in active state with $Z_{eff} \leq -1$ with respect to the time it spends poisoned with $Z_{eff} \geq 0$, when a daemon is not capable of capturing a new nucleus and exciting a scintillation. It would thus appear that (see Fig. 2) an evacuated detector should possess the highest efficiency, because the daemon, on having disintegrated the nucleus it had captured earlier, is not poisoned by nuclei of the air and, in approaching in the active state (with $-1 \geq Z_{eff} \geq Z$) the ZnS(Ag) layer, should practically with 100 per cent probability capture a nucleus here and excite a scintillation.

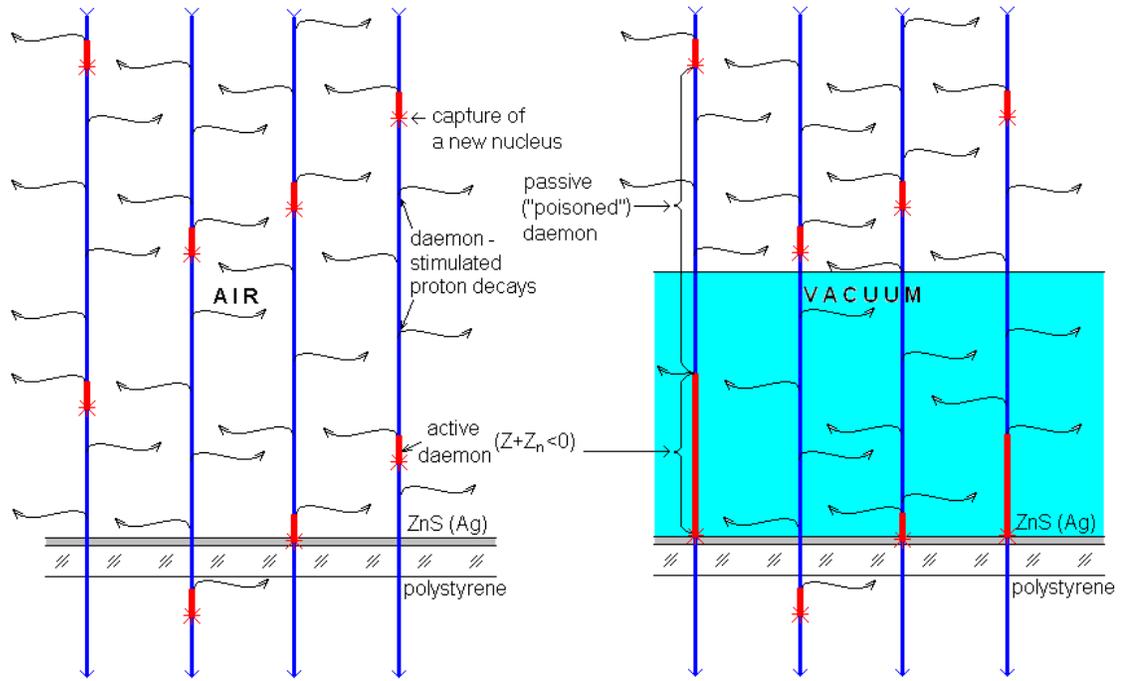

FIGURE 2. Positive effect of vacuum on the detection efficiency of the passage of negative daemons. Vacuum preserves the daemon in active state up to its entering the scintillator, because of preventing "poisoning" of the daemon by the excess positive charge of the nuclei captured by it from the air.

Considered from this standpoint, detectors filled with hydrogen or helium instead of air should possess a higher efficiency. Detectors filled with water vapor whose molecule has two protons could occupy an intermediate place. Because of a potential explosion hazard involved, we did not risk with starting experiments with hydrogen; experiments with helium cost us, however, the loss of two FEU-167 PM tubes (I.N.Kolyshkin warned us in time that helium diffusion through glass brings about deterioration of vacuum and PM tube degradation). Experiments performed in air with a humidity raised to 30-50% at $T = 28\text{-}30\ ^{\circ}\text{C}$ did not produce conclusive results. Nevertheless, the studies performed within the frame of the above considerations on possible effect of various factors on detection efficiency shed eventually light on some of the reasons for poor reproducibility of our results and pointed out certain ways to detector improvement.

## 3 THE MARCH 2005 EXPERIMENT AND ITS MAIN RESULTS

A new attempt at reproducing the results of March 2000 was undertaken by us in March 2005. The thickness of the ZnS(Ag) layer was increased by spraying to ~6 mg cm$^{-2}$, and the sensitivity of all PM tubes was again matched by properly adjusting their voltages. Besides, all the PM tubes we had at our disposal were classified by their intrinsic noise (at an output signal level of ≥2.5 mV). The lowest-noise tubes were mounted to view the bottom scintillators in the modules (the intrinsic noise signals, NLSs, of the upper PM tubes affect solely the trigger frequency of the system and not the final result, because we are processing only events with an HPS signal on the upper trace occurring at the trigger level of ≥2.4-2.6 mV, and the signals on the second trace are taken into account starting from the level ≥0.6 mV; recall that the PM tube anode load contains, besides an active 9.2-kΩ resistor, an inductance of 4.69 mH, through which the signal is supplied by a cable with 490-pF capacitance directly to the 30-pF oscillograph input; this L-C filter, while leaving the HPS amplitude unchanged, suppresses NLSs ~ three times). The thin (1 mm thick) top scintillator plates in modules #21-#24 were turned with the ZnS(Ag) layer up. Starting from the end of January 2005, the humidity in the air-conditioned room was maintained at a level of 30 per cent at $T = 28$ $^{\circ}$C.

The 8-module detector was exposed from the evening of March 5 to March 25, 2005 inclusive (483 h altogether). During this period, some parameters were varied as follows: (*i*) 15-20 March (116 h total time), the bottom PM tube of module #23 was screened with black paper; (*ii*) 16-19 March, the bottom PM tubes in modules #1 and #3 were interchanged (70 h; this time was not included in subsequent data treatment, but the experiment showed that the rate of the double event registration depends on the properties of the bottom PM tube). The observations yielded very interesting results.

All the eight modules could be clearly divided into three groups according to the number of recorded double events with HPSs triggering the top PM tube (events with |Δ$t$| ≤ 0.4 μs were rejected).

Three modules (#1, #2, and #24) recorded 62, 60, and 55 events each during this time, which is broadly in agreement with the count rate of double events in 2000 (when four modules, #1-#4, recorded from February 24 to March 27, 212 double events altogether with HPSs on the upper traces). The +20 < Δ$t$ < 40 μs bin of the $N(\Delta t)$ distribution contains 25 out of the total of 177 events (1.46$\sigma$ significance). But if normalizing the results to four modules and to the corresponding exposure time, we have to admit that we still have not reached the efficiency of the year 2000, i.e., some parameter remains uncontrollable. Nevertheless, adding these new data to those obtained in the months of March 2000, 2003, and 2004 (see Fig. 1) raises the significance of the 30-μs peak to 3.63$\sigma$ (C.L. = 99.97%).

Two of the modules (#21 and #22) recorded, accordingly, 103 and 119 events with a noticeable excess of events with Δ$t$ < 0 (57 and 69, respectively, compared with 46 and 50 for Δ$t$ > 0), which may be assigned to recording of an excess flux of objects propagating upward.

Finally, modules #3, #4, and #23 recorded a still larger number of double events. Module #3 recorded 959 events ($N(\Delta t)$ from -100 μs to +100 μs is 73, 96, 105, 85, 73, 101, 124, 102, 106, 94), module #4, 923 events (95, 98, 100, 97, 94, 87, 96, 90, 80, 86), and module #23, 436 events (42, 38, 42, 48, 52, 36, 44, 33, 61, 40) or, after subtraction of 116.5 h of exposure with black paper, 321 events (34, 25, 33, 35, 41, 22, 31, 20, 46, 34). If we restrict the signal amplitude in the upper channels to 3.5 ≤ $U_1$ ≤ 10.0 mV, and for the lower channels, to 0.6 ≤ $U_2$ ≤ 0.9 mV, then the number of such events in module #3 drops to 497 (40, 52, 55, 40, 32, 51, 68, 54, 64, 41), in module #4, to 613 (67, 63, 62, 68, 64, 53, 70, 56, 56, 54), and in module #23, to 230 (19, 14, 21, 25, 26, 20, 29, 18, 36, 22) [or to 168 (15, 9, 15, 18, 20, 12, 20, 13, 28, 18) if disregarding the events obtained with black paper], i.e.1340 events altogether. The total $N(\Delta t)$ distribution of these 1340 events is presented in Fig. 3. We readily see a 30-μs maximum with 167 events, which exceeds the average level by 33 events, standing out with a significance of $2.55\sigma$.[2]

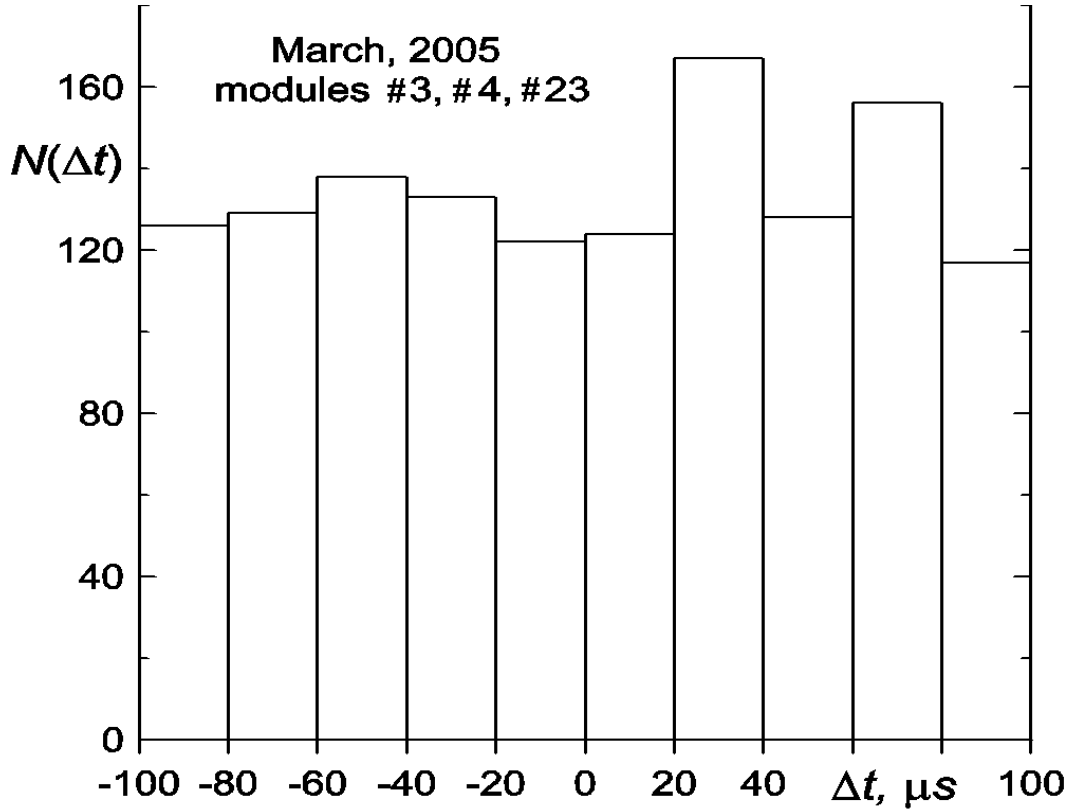

FIGURE 3. Total $N(\Delta t)$ distribution for March 5-25, 2005, of double events with HPS on the top scintillator obtained for modules #3, #4, and #23, where the bottom PM tubes have a thicker internal Al coating. The distribution was plotted for 1340 events with amplitude $3.5 \leq U_1 \leq 10.0$ mV in the upper channel and $0.6 \leq U_2 \leq 0.9$ mV in the bottom channel.

Calculations yield for the daemon flux corresponding to this excess (total area of the three modules 0.75 m$^2$, exposure time $1.75\times10^6$ s) $f_\oplus \approx 2.5\times10^{-9}$ cm$^{-2}$s$^{-1}$, a figure exceeding by a factor three to four the flux recorded in March, 2000 (18 excess events, total area of four modules 1 m$^2$, exposure time $2.5\times10^6$ s). It would seem that everything is all right, because the goal of the experiment has been reached, and we have demonstrated compatibility with the earlier results, so that the only thing still remaining to do was to eliminate some minor inadequacies.

## 4  ANALYSIS OF EXPERIMENTAL RESULTS. POSSIBLE ROLE OF THE INNER PM TUBE STRUCTURE

A question that immediately comes to mind is what could be the reason for such strong differences in behavior among different groups of modules having presumably identical parameters, including the PM sensitivity to light? Strange though as this might seem, an intrinsically noncontradictory answer can be put forward drawing again on the concept of the existence of daemons.

---

[2] It appears worth noting that the $N(\Delta t)$ distribution for double events with NLSs in the first channel of module #3 (with the total number of events about 2120) exhibits a maximum in the $-40 < \Delta t < -20$ μs bin with a significance $\approx 2.5\sigma$ (with no +30-μs maximum observed). Such features for NLSs were not seen earlier. We may consider it now to be of a purely stochastic nature.

We assumed that different PM tubes (or their different lots) exhibit, depending on the time and method of fabrication and the related inevitable deviations from the ideal standard technology, some features which, while not affecting markedly their spectral and photometric properties, give rise to different responses when crossed by daemons, as well as by components of other penetrating radiations.

As a matter of fact, we bought 12 FEU-167 PM tubes, part of which were selected for the first four modules, from Svetlana Co. (St.Petersburg) (located close to the Ioffe Institute) in 1998. It is at Svetlana Co. that the FEU-167 PM tubes with excellent characteristics had been developed in the early 1980s under G.S.Wildgrube. In the late 1980s, production of FEU-167 was transferred to Ekran Co. in Novosibirsk, where in June 2000 we bought four FEU-167 and 12 FEU-167-1 PM tubes. The best of them (as we believed) were used to replace the PM tubes in the already operated four-module detector which demonstrated poorer performance. Our PM tubes bought from Svetlana and Ekran had hardly the best performance, because they were chosen from the lots left over from what had been produced a few years before (for the well known reasons, industrial production in Russia had nearly stopped by the late 1990s).

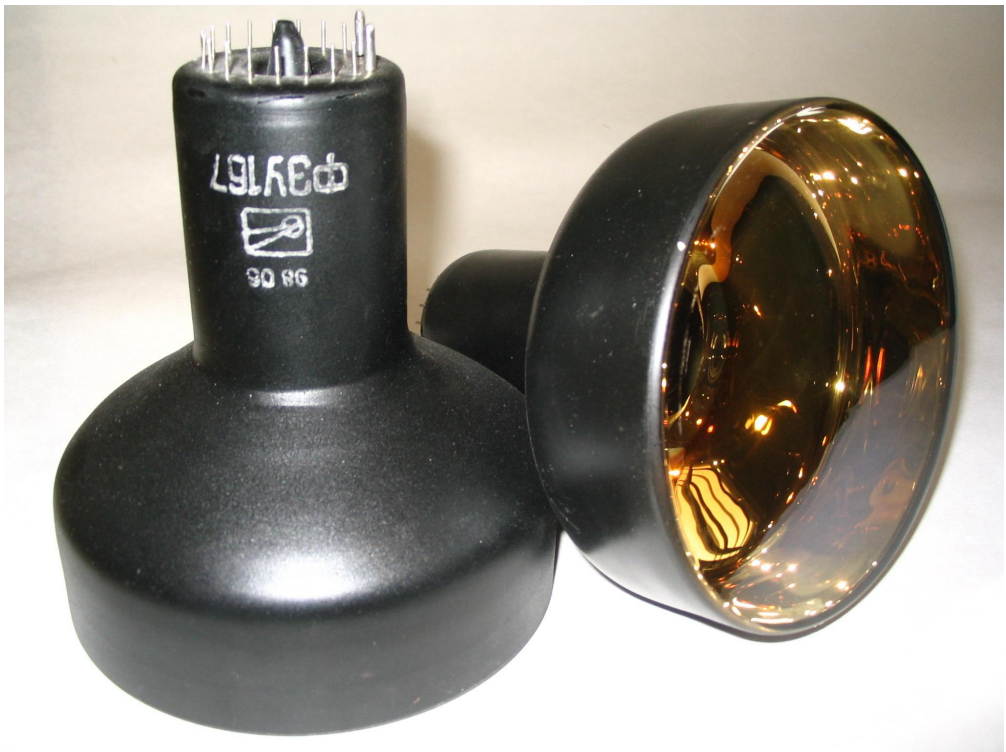

FIGURE 4. FEU-167 and FEU-167-1 PM tubes. Photo-cathode section diameter 125-130 mm.

To identify these unapparent parameters by which PM tubes could differ from one another, we had to study their design and technological details of their manufacture. The glass envelope of FEU-167 (see Fig. 4) consists of three parts: (*i*) a cylindrical photo-cathode section, 125-130 mm in diameter and about 40-45 mm long whose front part is protected by a glass disc $5_{-0.3}$ mm thick, (*ii*) a cylindrical stem, 50 mm in diameter and ~70 mm long, accommodating the dynodes and the anode, and (*iii*) a conical photo-cathode junction section 25-30 mm high. The envelope wall thickness is 2.3-2.5 mm. The photo-cathode and the junction sections are coated on the inside with a specular aluminum layer deposited by evaporation in vacuum. According to the original specifications of Svetlana, the thickness of the Al layer, although it is not strictly regulated because of not being able of influencing critically the PMT photometry properties, is about 0.1 µm. The surface of the Al layer and of the transparent front glass disc are coated by a transparent current-conducting Sb-Na-K-Cs photosensitive layer. The disc and the envelope of the photo-cathode section are made of S49-1 low-background boron-silicate (potassium-free) glass (67.5 wt.% $SiO_2$, 20.3 wt.% $B_2O_3$, 3.5 wt.% $Al_2O_3$, 8.7 wt.% $Na_2O$, Svetlana version) or S52-1 boron-potassium glass (Ekran version used in FEU-167-1, a PM tube similar in all other respects to FEU-167, $K_2O$ content 4.4 wt.%).

Based on the general idea of the capture, disintegration, and transport by a daemon of the remainder of the captured nuclei, we assumed that the incomplete reproducibility of the first results, as well as the above strong difference in the number of recorded double events in the three groups of modules, may be caused to a certain extent by differences in construction of different PM tubes too, in particular, by differences in thickness of the Al and photosensitive layer (referred to below as Al coating for brevity) in their photo-cathode section.

We carried out comparative measurements of the thickness of electrically conductive Al coating in different PM tubes. We used for this purpose two inductance coils (2 dia. 100-mm turns). On one of them the PM tube was placed with its front screen down, and the other was put on its conical section at a distance of 60 mm from the first one. A pulsed electric signal was supplied to the first coil, and the inductively induced pulse was obtained from the other. The system was calibrated by a set of Al foil discs, 140 mm in diameter and 0.05 µm thick, placed between the coils. The larger the total foil thickness, the weaker the inductive coupling between the coils, and the weaker the induced pulse, the sensitivity of the method decreasing naturally with increasing foil thickness. Because of the configurations of the foils (discs) and of the Al coating in a PM tube being different (gently slopping cone connecting with the cylinder), measurements of this type are not capable of yielding an absolute value of the thickness of the conductive coating in a PM tube (the more so that because of specific features of deposition it may turn out inhomogeneous in thickness). Nevertheless, they offer a judgment of the relative thickness of the current-conducting layer in different PM tubes.

The Svetlana-produced PM tubes were found to be, on the average, more noisy for output signals ≥2.5 mV than their Novosibirsk counterparts, and therefore, in our new arrangement all the lower PM tubes in all the modules turned out to be Novosibirsk-made (FEU-167 in module #1, and the FEU-167-1 in others). The thickness of the Al coating in the Novosibirsk PM tubes turned out, however, to be ~3-10 times that of the Svetlana devices. Moreover, it was found that the bottom PM tubes in modules #3, #4, and #23 belong to a group with the largest coating thickness.

## 5  PM TUBE AS A VACUUM DETECTOR OF DAEMONS

The above adds up to the following, inevitably strongly simplified scenario of daemon passage through the modules containing PM tubes with internal Al coatings of different thickness and with envelopes made of different glass (low-background, i.e., potassium-free, or boron-potassium).

By our standard scenario (see Fig. 2), if a daemon propagates through air, it becomes poisoned, as it were, by the N or O nuclei it captures, for a long time, because, on having acquired an excess positive charge, it is no longer capable of capturing and exciting a new nucleus. Having "digested" in 5-10 µs the excess protons and assumed $Z_{eff} = |Z_n| - |Z| < 0$, the daemon becomes active for a certain time and, with $V \approx 10\text{-}15$ km s$^{-1}$, propagates through air ~1-4 mm until the capture of another nucleus (Drobyshevski *et al*., 2003a). If during this interval it crosses the ~10-µm-thick ZnS(Ag) layer, it will capture with a high probability a nucleus here and excite it to produce an HPS. (Recall that the passage of a daemon or of cosmic ray particles through the upper PM tube generates only NLSs, so that these events are not considered here.) Continuing its downward flight, the daemon, still being poisoned by the remainder of the nucleus, at $V > 5$ km s$^{-1}$ crosses the bottom scintillator without producing a scintillation. The bottom scintillator will be excited if by the time it approaches the bottom tinned-iron lid of the casing the daemon becomes again active and captures a nucleus here. The numerous electrons emitted in the capture cross the 22-cm-layer of air separating them from the bottom ZnS(Ag) scintillator, and excite here an NLS; and it is this that causes the double event. This is the scenario of ideal detector operation, which apparently was primarily realized in March 2000, when the Svetlana PM tubes were used.

But if, on crossing the upper ZnS(Ag) layer and exciting there an HPS, the daemon reaches the level of the lower lid of the casing and enters the PM tube, it may, while passing the 4-5-cm path in vacuum, lower its $Z_{eff}$ to $\approx -(4 \div 6)$ as protons in the remainder of the captured nucleus continue to disintegrate.

Obviously enough, the response of a PM tube to the passage of such a daemon should depend critically on structural details of its photo-cathode section.

If the thickness of the Al coating inside it is $\leq 0.1$ µm, the daemon with $Z_{eff} \approx -(4 \div 6)$ will cross the electrically conductive coating without interaction with its nuclei, enter the glass and, only on capturing here a B, Na, Si, or Al nucleus (S49-1 glass) with ejection of a small number of electrons (10-15) contained in the nearest (dielectric) environment of the excited nucleus, will generate in the PM tube a weak NLS. Note that the number of secondary electrons is small, because the work function of an electron from an inorganic dielectric is comparable to the ionization energy of its constituent molecules (~10 eV). If, however, the thickness of the Al coating inside the PM tube is $\geq 0.5$ µm thick, the daemon with $Z_{eff} \approx -(4 \div 6)$ propagating with $V = 10$-15 km s$^{-1}$ will capture here a nucleus with a practically 100% efficiency (recall that calculations similar to those performed by Drobyshevski *et al*., (2003a) yield for Al $\lambda \times |Z_{eff}| \approx 2.5$ µm for $V = 10$-15 km s$^{-1}$ ($\lambda$ is the mean free path to nucleus capture). In addition to 13 atomic electrons, the nucleus excited in the capture will eject out of the Al layer a large number of refilling electrons (in the metallic phase), as well as of secondary electrons, because the energy needed to release an electron here (the work function), in contrast to that in the glass, is low. For pure Al it is ~4 eV, and for Al coated by thin layers of Na, K, Cs, Sb and their compounds, ~1 eV.

This is why a PM tube with a thick enough internal Al mirror is capable of operating by itself, without a scintillator, as an efficient vacuum detector of daemons (with an efficiency as high as tens of per cent).[3] It is clear also that because of avalanche multiplication of electrons in a thick Al film such a PM tube will respond to and detect cosmic rays and intrinsic radioactive radiations crossing it with a higher efficiency than a tube with a thin Al layer. This is why the number of recorded stochastic double events (background) in modules #3, #4, and #23 is very large. We may recall also that the envelopes of the bottom FEU-167-1 PM tubes are made of potassium-containing glass, so that a sizable fraction of small-amplitude NLSs triggered by them may in actual fact be initiated by the β radioactivity of $^{40}$K. It appears appropriate to point out here that cosmic rays, which produce double NLSs with $\Delta t = 0$, on the average, once every few minutes, and sometimes simultaneously in several modules, generate these signals in the PM tubes themselves (including those screened by black paper); judging from the rate of their appearance, such double NLSs with $\Delta t = 0$ ensue from cosmic rays in the dynode part of a PM tube.

---

[3] The idea is in no way new. Application of PMT (more exactly, of its dynode assembly) as an electron multiplier for direct high-efficiency particle detection is nearly seven decades old (e.g. Lifshits, 1953; Price, 1958).

The response of PM tubes with an intermediate Al mirror thickness to daemons (and background radiations) should naturally be more complex than that with a thin or thick Al layer. It will depend in a large measure on parameters of the electron optics governing the motion of electrons with different energies and emitted from different parts of internal surface of the photo-cathode section, as well as on the composition of the envelope and disc glass.

It is along these lines that one may look for explanation of the fact that the bottom PM tubes in modules #21 and #22 detect preferentially daemons arriving from below (the excess of events with $\Delta t < 0$). It may be conjectured that it is primarily the low-energy electrons emitted in the capture of nuclei by daemons in K-containing S52-1 glass of the disc that are intercepted by the dynode section with their subsequent multiplication than the ones ejected from the Al-coated walls of the conic section or side cylindrical walls. The role of secondary electron emission produced as electrons strike the glass walls and the comparatively thin Al coating is unclear (one should not forget that all these surfaces are coated by a thin (~10 nm) Sb-Na-K-Cs active layer).

While the efficiency of detection of daemon passage with $V \sim 10$ km s$^{-1}$ through a ZnS(Ag) layer can be estimated as a ratio of a few mm of air path in active state to ~10 cm of flight in poisoned state (see Fig. 2), i.e., as a few (~3) per cent (so that our detector made up of two such layers has an efficiency of ~$10^{-3}$),[4] the efficiency with which a daemon crossing a PM tube with a thick Al mirror could be detected may amount to tens of per cent. Assuming the area of the PM tube front disc to be 115 cm$^2$, and accepting that the 30-μs maximum in the $N(\Delta t)$ distribution in Fig. 3 exceeds the background during 20 days by 33 events, and assuming also that the bottom PM tubes in modules #3, #4, and #23 are identical, we obtain that the flux of daemons with $V \sim 10$-15 km s$^{-1}$ measured in March 2005 is $f_\oplus \sim 0.5 \times 10^{-7}$ cm$^{-2}$ s$^{-1}$. This figure exceeds ~50 times the flux measured by our detector earlier (Drobyshevski, 2002).[5] Allowing for the low efficiency of the top scintillator-PMT assembly (~3%), we arrive at $f_\oplus \sim 10^{-6}$ cm$^{-2}$s$^{-1}$ for the real flux. This value is consistent with our earlier estimates of the daemon flux ~$3 \times 10^{-7}$ cm$^{-2}$s$^{-1}$ from heliocentric orbits (Drobyshevski, 1997) and with a value of $10^{-7}$-$10^{-6}$ cm$^{-2}$s$^{-1}$ derived from analysis of the origin and behavior of the "Troitsk anomaly" observed near the end-point of the tritium β spectrum obtained in neutrino mass measurements (Drobyshevski, 2005a).

An additional argument in favor of the working hypothesis that it is a PM tube itself that responds to daemons is offered by the experiment with module #23, in which the bottom PM tube scintillator was screened during 116.5 h by black paper (from 18$^h$20', 15.03.2005 to 20$^h$00', 20.03.2005). Strange as this might seem, the significance of the 30-μs maximum in experiment with the PM tube screened during nearly 5 days, rather than dropping, increased even somewhat. The statistics involved here is naturally poor, so that further experiments with PM tubes screened from the scintillators are under way in the Baksan Neutrino Observatory, yielding support for this view (Drobyshevski, Kuzminov and Drobyshevski, in preparation).

---

[4] The low efficiency is suggested also by the fact that a year-long exposure in 2004 of modules #1-#4 mounted under modules #21-#24 (and in the opposite arrangement) no reliable triple event involving shifted HPSs on the top scintillators in the modules arranged one over another was recorded.

[5] On the other hand, taking into account the sensitivity of a PM tube itself to the passage of a daemon we have discovered, one should not overlook the possibility that a part of the NLSs on the bottom PM tubes detected in previous experiments were also generated by the passage of daemons through them rather than by scintillations in the bottom ZnS(Ag) layer induced by electrons from the lower lid of the casing which were emitted by the nucleus captured here by the daemon, as we had been inclined to believe before. Further studies are obviously needed to clarify this point.

# 6  MAIN CONCLUSIONS AND PROSPECTS

The March, 2005 experiments have yielded interesting results in two main directions, astrophysical and methodological.

Starting with the *astrophysics*, it has been shown that:

(1)    The March maximum of the low-velocity ($V \sim$ 10-15 km s$^{-1}$) ground-level daemon flux incident from NEACHOs is apparently far in excess of the lower limit $f_\oplus \sim 10^{-9}$ cm$^{-2}$s$^{-1}$ set by previous experiments and reaches $f_\oplus \sim 0.5 \times 10^{-7}$ cm$^{-2}$s$^{-1}$, a figure approaching our old estimates drawn on theoretical grounds (Drobyshevski, 1997).

(2)    Summing up the data obtained in 2000, 2003, 2004, and 2005 raises the significance of the 30-µs March maximum in the $N(\Delta t)$ distribution to 3.63$\sigma$ (to 99.97%; see Fig. 1).[6]

Now to *methodology*:

(1)    We have developed an understanding that our two-scintillator detector with thin ZnS(Ag) layers has a fairly low sensitivity ($\sim 10^{-3}$) because of the small ratio of the times the daemon moving through air spends in the active state, in which it can capture a new nucleus, to that in passive state, where it is "poisoned" by nuclei with $|Z_n| \geq |Z|$.

(2)    This understanding has helped us to realize the merits of evacuated detectors, where the daemon which has reached active state by "digesting" protons in the captured nucleus would reside in this state until it enters the sensitive (e.g., scintillator) layer.

(3)    The experiment has revealed that some FEU-167-1 PM tubes, which by their spectral and photometric characteristics do not stand out among other FEU-167s, possess the properties of a high-efficiency evacuated daemon detector. Inductive measurements of the thickness of the electrically conducting coating inside the cylindrical cathode section (dia. 125 mm, ≈45 mm long) showed it to exceed the specified value ($\sim$0.1 µm) by a few times to reach $\sim$1 µm. It is conceivable that the daemons that have become active after entering the evacuated cathode section of the PM tube retain and even increase it (the value of $|Z| - |Z_n| > 0$ increases) in crossing the 45-mm-long section and, on striking the fairly thick layer of electrically conducting material (Sb-Na-K-Cs photosensitive layer and Al coating), capture with a high probability a nucleus here, which gives rise to emission of hundreds of electrons (internal conversion, Auger, refilling, and secondary) generating a noticeable PM tube signal. Capture of a nucleus in the thin layer of deposited aluminum or in the Sb-Na-K-Cs coating should occur with a very low probability and, if taking place in glass, would bring about emission of only a few tens of electrons (there would be very few secondary and refilling electrons).

Thus, in addition to the urgently needed experiments on PM tubes with screened photo-cathodes (front glass discs), one could conceive of at least two areas where detection of daemons could be improved methodologically so as to raise its efficiency:

(a) Placing thin-layer scintillator detectors in vacuum.

(b) Using purposefully modified PM tubes, more specifically, (*i*) with a larger thickness and/or variable composition of the electrically conducting cathode coating, (*ii*) with a longer cathode section to increase the fraction of daemons that become active here and the degree of their activity, i.e., the difference $|Z| - |Z_n| > 0$, and (*iii*) using blind PM tubes (if such devices could still be called PM tubes; they are actually electron multipliers (Lifshits, 1953)) with front discs also coated on the inside by, possibly, opaque to light, electrically conducting layers with optimally chosen thickness and atomic composition. One cannot rule out the possibility that the composition of the envelope and disc glass could also be of importance here.

---

[6]  Taking into account data of modules #3, #4, and #23 would naturally increase still more the significance of the 30-µs peak, but these results were obtained, in a sense, with another, much more efficient detector, whose parameters have yet to be appraised and quantified. Therefore it would hardly be possible to carry out presently indisputably correct summation of confidence levels of these two data sets.

Although the confidence level of the main results obtained in this study is in itself not very high (~90-95%), nevertheless, they suggest new prospects and directions for further investigations and use of the interesting information we have gained. Moreover, adding the new data to what had been amassed in previous experiments raises the confidence level of the existence of the March maximum in the daemon flux from NEACHOs to 99.99%, let alone the weight it renders to the argument that the totality of observations collected by us thus far defies any other interpretation except the scenario proposed within the daemon concept, the basis underlying our experiments.

## ACKNOWLEDGMENTS

The material exposed in the paper has been reported and discussed at a Seminar at the Ioffe Physico-Technical Institute on November 14, 2005.

I am thankful to N.N.Nikonov for draft processing of experimental data. I express also my gratitude to Dr E.A.Bogomolov for stimulating discussions and for the many-month-long, thus far unavailing, search for an interpretation, other than daemon based, of the experimental data obtained in this purposeful study (as a minor note, I cannot agree with him that in a calculation of statistical significance of the maxima in the histograms their deviation from the weighted mean should be preferably divided by the arithmetic sum of errors rather than by their square sum).

# REFERENCES

Bahcall J. H., Flynn C. and Gould A. (1992). Local dark matter from a carefully selected sample. *Ap. J.*, **389**, 234-250.

Barrow J. D., Copeland E. J. and Liddle A. R. (1992). The cosmology of black hole relics. *Phys. Rev. D*, **46**, 645-657.

Drobyshevski E. M. (1996). Solar neutrinos and dark matter: cosmions, CHAMPs or... DAEMONs? *Mon. Not. R. Astron. Soc.*, **282**, 211-217.

Drobyshevski E. M. (1997). If the Dark Matter objects are electrically multiply charged: New opportunities. In: Klapdor-Kleingrothaus H. V. and Ramachers Y. (Eds.), *Dark Matter in Astro- and Particle Physics*, World Scientific, Singapore, pp. 417-424.

Drobyshevski E. M. (2000). Time of explosive decay of daemon-containing nucleus. *Mon. Not. R. Astron. Soc.*, **311**, L1-L3.

Drobyshevski E. M. (2002). Detecting the dark electric matter objects (daemons). *A&ATrans*, **21**, 65-73.

Drobyshevski E. M. (2005a). Daemons, the "Troitsk anomaly" in tritium beta spectrum, and the KATRIN experiment. Preprint (*hep-ph*/0502056).

Drobyshevski E. M. (2005b). Detection of Dark Electric Matter Objects falling out from Earth-crossing orbits. In: Spooner N. J. C. and Kudryavtsev V. (Eds.), *Proc. of the 5th Intnl. Workshop on "The Identification of Dark Matter"*, World Scientific, Singapore, pp. 408-413.

Drobyshevski E. M., Beloborodyy, M. V., Kurakin R. O., Latypov V. G. and Pelepelin K. A. (2003a). Detection of several daemon populations in Earth-crossing orbits. *A&ATrans*, **22**, 19-32.

Drobyshevski E. M., Drobyshevski M. E., Izmodenova T. Yu. and Telnov D. S. (2003b). Two years of detecting DM objects – the Solar system members. *A&ATrans*, **22**, 263-271.

Ivanov P., Naselsky P. and Novikov I. (1994). Inflation and primordial black holes as dark matter. *Phys. Rev. D*, **50**, 7173-7178.

Lifshits T. M. (1953). Using electronic multipliers for count of elementary particles and quanta. *Sov. Phys. – Uspekhi*, **50**, 365-432.

Markov M. A. (1966). Elementary particles with largest possible masses (quarks and maximons). *ZhETF*, **51**, 878-890.

Price W. J. (1958). *Nuclear Radiation Detection*, McGraw-Hill Book Co, N.-Y. – Toronto – London.

Taylor J. R. (1982). *An Introduction to Error Analysis*, Univ. Science Books, Mill Valley, CA.